\documentclass[12pt]{article}

\def\centerbmp#1#2#3{\vskip#2\relax\centerline{\hbox to#1{\special
  {bmp:#3 x=#1, y=#2}\hfil}}}

\usepackage{latexsym}
\usepackage{epsfig}
\usepackage{setspace}
\usepackage{graphicx}
\usepackage{natbib}
\usepackage[latin2]{inputenc}
\begin{document}

\title{Fast and Exact Sequence Alignment with the Smith-Waterman Algorithm: The SwissAlign Webserver}
\author{Gábor Iván\,$^{\rm a,b}$ \and  Dániel Bánky\,$^{\rm a,b}$\\ \and Vince Grolmusz\,$^{\rm a,b}$\\ \\
$^{\rm a}$PIT Bioinformatics Group\\ Eötvös University, H-1117 Budapest, Hungary\\ $^{\rm b}$Uratim Ltd. H-1118 Budapest, Hungary}

\date{}

\maketitle

\begin{abstract}
It is demonstrated earlier that the exact Smith-Waterman algorithm yields more accurate results than the members of the heuristic BLAST family of algorithms. Unfortunately, the Smith-Waterman algorithm is much slower than the BLAST and its clones.
Here we present a technique and a webserver that uses the exact Smith-Waterman algorithm, and it is approximately as fast as the BLAST algorithm. The technique unites earlier methods of extensive preprocessing of the target sequence database, and CPU-specific coding of the Smith-Waterman algorithm.
The SwissAlign webserver is available at the http://swissalign.pitgroup.org address.
\end{abstract}

\section{Introduction}

The BLAST (Basic Local Alignment Search Tool) algorithm \cite{Altschul1990} and its versions are among the most important algorithmic applications in biology. The speed of BLAST made it a much more frequently used algorithm, than the exact Smith-Waterman algorithm for sequence alignments \cite{Smith1981}. Most of the users of the BLAST algorithm are not aware of that for the speed they need to pay with accuracy: extensive studies witnessed that the Smith-Waterman algorithm finds alignments that are overlooked by BLAST \cite{Pearson1991, Pearson1995, Shpaer1996}.

Several methods were reported to speed up the slow, but exact and reliable Smith-Waterman algorithm \cite{Smith1981} with artificial intelligence applications \cite{Eddy2011}, with parallelization \cite{Bandyopadhyay2009, Hasan2011}, with CPU-specific command-sets \cite{Rognes2000, Farrar2007, Rognes2011, Szalkowski2008}, clustering \cite{Itoh2004}, preprocessing \cite{Itoh2005}, or advanced heuristic approaches \cite{Smith2006}, some of these improved the Smith-Waterman implementations to achieve a speed comparable to that of the BLAST variants \cite{Farrar2007}. All of these methods are important since they make the exact Smith-Waterman algorithm applicable for large datasets in reasonable time.

In numerous sequence alignment tasks, one sequence, the \emph{query sequence}, is aligned to a database of sequences; then the top-scoring hits are returned, showing their alignment to the query sequence. 

Our aim was to provide an exact and reliable alternative to the sequence alignment web servers utilizing BLAST and its clones. The main idea behind BLAST is to apply heuristics when calculating local alignments; the main assumption is that a high-scoring local alignment will include a substring of three consecutive amino acids or nucleotides from each sequence that achieve a high score when aligned without gaps \cite{Altschul1990}. The motivation behind using such heuristics is the possible gain in speed: the BLAST-related algorithms perform significantly faster than the Smith-Waterman algorithm \cite{Altschul1990, Rognes2000, Farrar2007, Rognes2011}. 

The speed of query-processing can be improved even further if the query sequence is chosen from the target database itself. In this case, all the information is known \emph{a priori}, therefore one can pre-compute and cache the score of the highest-scoring local alignment for each sequence pair, however, it is enough to store scores for sequence pairs that achieve a statistically significant score, characterized by an e-value limit.

To demonstrate what is possible when using an improved Smith-Waterman algorithm and caching the pre-processed values, we built a web server that accepts sequences from the Swiss-Prot subset of the UniProt database \cite{Consortium2010} as queries, and returns the top alignments of the Swiss-Prot DB to that query sequence. The results are usually available within one second with default settings.
\vskip -0.5cm

\section{Results and discussion}

The SwissProt subset of the UniProt database \cite{Consortium2010} contains the reviewed subset of the whole UniProt database. The size of the whole UniProt database, that is, the disjoint union of Swiss-Prot and TrEMBL, is close to 24 million entries. The Swiss-Prot database contains 536,789 sequences on August17, 2012.  

The SwissAlign webserver applies extensive preprocessing, as suggested in \cite{Itoh2004, Itoh2005} and CPU-specific instruction-set \cite{Farrar2007} for the alignment score computation. 

More exactly, our web server does
\begin{enumerate}
\vskip -1cm
\item Offline Calculation of scores for each sequence pair from the input database, in our case the Swiss-Prot DB. Only the scores of sequence pairs above a given e-value limit (in our case, $10^{-3}$) are stored.
\item Cache the pre-calculated scores in a database. The database is queried whenever a sequence is input in the webserver; only the top-scoring sequences are returned.
\item Aligning the query sequence to all the top scoring sequences returned in the previous step and displaying the alignments in plain text and HTML format.
\end{enumerate}

\begin{figure*}
\centering
\includegraphics[width=14cm, height=6cm]{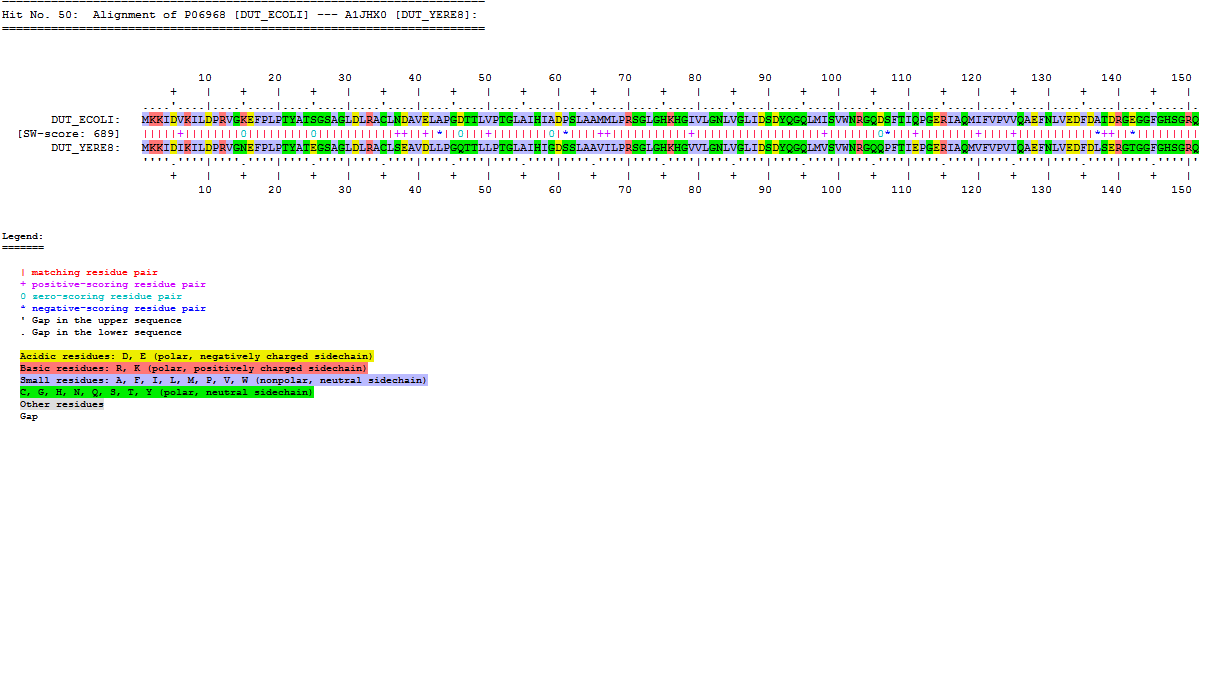}
\caption{\small \sl Sample output of the SwissAlign server.}
\end{figure*}

First we pre-computed and stored all the local alignment scores above the e-value limit of $10^{-3}$. Then, after this expensive step, we only need to update every week our alignment-score database, when Swiss-Prot gets new entries.

The algorithm used for local alignment score calculation is the one described in \cite{Farrar2007}. It uses hardware-specific optimizations that make the Smith-Waterman algorithm at least an order of magnitude faster than the "naive" implementation. Improvement in processing speed is achieved by working on multiple data in parallel using the SSE2 instruction set available in x86 CPUs, and also preprocessing the query sequence and arranging the data in RAM so that memory accesses occur in an optimized pattern. We encapsulated this algorithm in a program that performs incremental update of the database, adding the scores of the missing sequence pairs. Note, that constructing the database the first time is equivalent to an incremental update performed on a zero-sized database.

{\bf The Application of the Webserver:} The webserver finds the best alignments to any query-sequence from the Swiss-Prot database in the Swiss-Prot database. The query sequence can be specified by its accession number \cite{Consortium2010};  a link is provided next to the input field for comfortable accession number lookup.

Next, one may leave the NCBI taxonomy ID field blank. In this case, the sequence belonging to the entered Swiss-Prot accession number will be aligned to all the sequences in Swiss-Prot; alignments of top-scoring sequences will be returned. The number of the returned sequences can be adjusted from 3 through 500; the processing time is proportional to this number. One may also provide a valid NCBI taxonomy identifier in the NCBI taxonomy ID field, while not checking the "Include subtree" field. In this case, the entered Swiss-Prot sequence will be aligned to all the Swiss-Prot sequences belonging to the entered NCBI taxonomy identifier. For example, typing '562' in the NCBI taxonomy ID field filters the output sequences to contain only alignments of sequences belonging to NCBI taxonomy ID 562 ({\em Escherichia coli}). 

In the previous example, a significant number of sequences belonging to {\em Escherichia coli} may not be present in the output, because they have a finer taxonomy classification, therefore they have different NCBI taxonomy identifiers. If all the sequences having taxonomy identifiers related to {\em Escherichia coli} are required in the output, the "Include subtree field" has to be checked (along with providing the NCBI taxonomy ID of {\em Escherichia coli} at the NCBI taxonomy ID field). 

The input of the NCBI Taxonomy ID is made comfortable with an autocompletion feature.

Query results, i.e., the top-scoring alignments to the query sequence, are presented in text and HTML format.
In the HTML format output, amino acids of the aligned sequences are coloured based on their basic chemical properties (see Fig. 1), the output of the webserver contains a legend for each color and each symbol used. 
An instruction video is also available at http://swissalign.pitgroup.org/swissalign.swf.

{\bf Conclusions:} We demonstrated a web server that aligns a selected sequence of the Swiss-Prot database to the remaining sequences of Swiss-Prot exceptionally fast. The alignments are obtained using a non-heuristic method: an optimized version of the Smith-Waterman algorithm. The target sequences can be filtered by their NCBI taxonomy identifier, and the results are displayed in text and coloured HTML format. 
\vskip -0.6cm


\end{document}